\documentclass[twocolumn,prd,preprintnumbers,nofootinbib,showpacs]{revtex4}

\usepackage{graphicx}
\usepackage{bm}
\usepackage{amssymb,amsmath}

\usepackage{array}

\def\beqn{\begin{eqnarray}}
\def\eeqn{\end{eqnarray}}
\def\barr{\begin{array}}
\def\earr{\end{array}}
\def\btab{\begin{tabular}}
\def\etab{\end{tabular}}
\def\bite{\begin{itemize}}
\def\eite{\end{itemize}}
\def\bcen{\begin{center}}
\def\ecen{\end{center}}

\def\eq{\begin{equation}}
\def\ee{\end{equation}}
\def\nn{\nonumber}

\def\kdagger{K\hspace{-0.3cm}/}

\def\pgdagger{P\hspace{-0.3cm}/}

\def\keldagger{k\hspace{-0.2cm}/}

\def\q2dagger{q_2\hspace{-0.35cm}/\;}

\begin{document}


\title{Dispersive contributions to $e^+p/e^-p$ cross section ratio in forward regime}

\author{Mikhail Gorchtein} 
\affiliation{
California Institute of Technology, 
Pasadena, CA 91125, USA} 

\date{\today}

\begin{abstract}
Two-photon exchange (TPE) contributions to elastic electron-proton scattering 
in the forward regime are considered. The imaginary part of TPE amplitude 
in these kinematics is related to the DIS nucleon structure functions. The 
real part of the TPE amplitude is obtained from the imaginary part by means of 
dispersion relations. We demonstrate that the dispersion integrals for the 
relevant elastic $ep$-scattering amplitude converge and do not need 
subtraction. This allows us to make clean prediction for the real part of the 
TPE amplitude at forward angles. 
We furthermore compare $e^+p$ and $e^-p$ cross sections which depends on 
the real part of TPE amplitude, and predict the positron cross section to 
exceed the electron one by a few per cent, with the difference ranging from 
1.4\% to 2.8\% for electron $lab$ energies in the range from 3 to 45 GeV. 
We furthermore predict that the absolute value of this asymmetry grows with 
energy, which makes it promising for experimental tests.

\end{abstract}

\pacs{12.40.Nn, 13.40.Gp, 13.60.Fz, 13.60.Hb, 14.20.Dh}

\maketitle


\section{Introduction}
\label{sec:intro}
Recently, much attention has been attracted by the two-photon exchange (TPE) 
contribution to the elastic electron-proton scattering. On the one hand, the 
recent experimental data on $G_E$ to $G_M$ ratio at higher momentum 
transfers show significant discrepancy between the results obtained with the 
new polarization transfer technique \cite{gegm_pol} and those obtained using  
Rosenbluth separation \cite{gegm_ros}. Since the TPE contribution is the only 
largely unknown order $\alpha_{em}$ correction to the elastic $ep$-scattering, 
it was argued that a proper inclusion of these effects into the analysis of 
the $ep$-scattering observables would reconcile the two measurements 
\cite{marcguichon}. 
Recent theoretical calculations of the real part of the 
TPE amplitude in different models seem to support this idea, 
though at the qualitative level \cite{cs_theo1}, \cite{cs_theo2}, 
\cite{cs_theo3}. At the moment, it is only the 
elastic (nucleon) intermediate state contribution together with the associated 
with it IR divergent part, needed for an analysis of the experimental data, 
that one can be confident about \cite{cs_theo1}. 
The partonic model ``handbag'' calculation of Ref. \cite{cs_theo3}, though 
definitely represents important physics contribution at high energy and 
momentum transfer, suffers of unphysical IR divergencies which can be 
presumably cancelled by including the ``cat ears'' mechanism. 
As for the further inelastic contributions, only estimates with the 
$\Delta(1232)$ for the real part of the TPE amplitude exist in the literature 
\cite{cs_theo2}.

\indent
Another experimental test of TPE effects is proposed at JLab \cite{epm_jlab} 
and at VEPP-3 storage ring \cite{epm_vepp}, 
where the $e^+p/e^-p$ cross section ratio will be studied in a wide angular 
range. A deviation from 1 originates to leading order in $\alpha_{em}$ from 
the interference between the one photon exchange contribution which is 
linear  in lepton charge, and the real part of the TPE contribution, 
which is quadratic in lepton charge. 

\indent
On the other hand, over past 5 years new observables become accessible 
experimentally, the single spin asymmetries with beam electron or target 
(recoil) proton polarized normally to the reaction plane \cite{bn_exp}. 
These observables have been first studied theoretically back 
in 1970's \cite{derujula} and were shown to be directly sensitive to the 
imaginary part of the TPE amplitude. Furthermore, all IR divergent terms cancel 
in these asymmetries (unlike the cross section), 
which makes them especially attractive since they 
offer the most clean experimental tests of the TPE effects. 
There is a considerable interest to this class of observables from the theory 
point of view, as well \cite{bn_theo}.

\indent
The TPE amplitude is related to Compton scattering with two space-like photons 
(virtual-to-virtual Compton scattering - VVCS) which is largely unknown. 
For the inelastic intermediate states contributions, it is 
only the imaginary part of the VVCS amplitude in the forward direction 
that is well studied, since it can be expressed in terms of the DIS 
structure functions. So, the only kinematical point where a firm prediction for 
the TPE amplitude can be made is the exact forward limit. Unfortunately, all 
the observables which measure the TPE effects necessarily vanish at that point. 
It was proposed \cite{afanas} to combine the DIS input in forward direction 
with the phenomenological $t$-dependence taken from that of the Compton 
scattering differential cross section $\frac{d\sigma}{dt}$ which is measured at 
high energies and low values of $t$. 
Modelling in this way the imaginary parts of the electron 
helicity-flipping $ep$-scattering amplitudes, it was possible to successfully 
describe the data on beam normal spin asymmetry at forward angles 
\cite{javvcs}. The motivation of this work is to extend the 
phenomenological approach of the previous paper and compute the imaginary parts 
of the electron helicity-conserving amplitudes. The real parts which enter the 
expression for the elastic cross section and the charge asymmetry are obtained 
through dispersion relations.
Combined with the phenomenological $t$-dependence which proved itself plausible 
for the imaginary part, it will allow us to predict the high energy dispersive 
contributions to the $e^+p/e^-p$ cross section ratio and the elactic cross 
section. 

\indent
For the case of low electron energies, a dispersive approach was adopted in 
\cite{bernabeu}. The authors of that work used once-subtracted dispersion 
relations in the annihilation channel $t$ at fixed energy $\approx0$. At this 
energy, the hadronic part of the TPE graph may be parametrized through the 
nucleon polarizabilities. To calculate the subtraction constant, the authors 
notice that the imaginary part of the hadronic amplitude at $t=0$ is given 
in terms of DIS structure functions. They furthermore use dispersion relations 
for these structure functions, and by doing the low energy expansion express 
the corresponding real parts as moments of the structure functions. After that, 
these real parts are embedded into the TPE amplitude. 

\indent
Instead, our approach is designed for high electron energies, therefore no 
connection 
with the polarizabilities can be made, neither the low energy expansion can be 
performed to express the final result as an integral over the DIS structure 
functions' momenta. First, we calculate the imaginary part of the TPE 
amplitude in terms of the electron phase space integral over DIS structure 
functions and obtain the imaginary parts of the invariant $ep$-scattering 
amplitudes. Instead of using dispersion relations for the hadronic amplitude 
separately, we proceed with the dispersion relations for the invariant 
amplitudes for elastic $ep$-scattering and demonstrate that they converge and 
no subtraction is needed. In this way, we obtain a clean prediction for the 
real parts of these amplitudes and the observables which they enter. 

\indent
The imaginary parts of helicity-flip amplitudes were shown to be dominated by 
the region of low virtualities of the exchanged photons, and it prooved 
adequate to approximate the DIS structure function $W_1$ by its value at the 
real photon point, where it is related to the total photoabsorption cross 
section. However, this approach is not expected to work in the case of the 
helicity-conserving amplitudes, especially for their real parts where all the 
values of these virtualities are allowed. Therefore, we include in our 
calculation the phenomenological form of the $Q^2$ dependence of the virtual 
photon cross section obtained from low-$x$ data. 
For the moment, we will concentrate on the observables averaged over the 
nucleon spin, i.e. the cross section and the beam normal spin asymmetry. 
We leave the consideration of the target normal spin asymmetry to an upcoming 
work.

The paper is organized as follows. In Section \ref{sec:el_ampl}, we overview 
the elastic $ep$-scattering amplitude and the kinematics of the reaction. 
The calculation of the imaginary parts of the invariant $ep$-scattering 
amplitudes is performed in Section \ref{sec:im_2gamma}. These imaginary parts 
are then used as input for the dispersion relations to determine their real 
parts in Section \ref{sec:re_2gamma}, and the convergence of these dispersion 
relations is discussed. In Section \ref{sec:results}, we present the results 
of the calculation.

\section{Elastic $ep$-scattering amplitude}
\label{sec:el_ampl}
In this work, we consider elastic electron-proton scattering process 
$e(k)+p(p)\to e(k')+p(p')$ for which we define:
\beqn
P&=&\frac{p+p'}{2}\nn\\
K&=&\frac{k+k'}{2}\nn\\
q&=&k-k'\;=\;p'-p,
\eeqn
and choose the invariants $t=q^2<0$ \footnote{In elastic $ep$-scattering, the 
usual notation for the momentum transfer is $Q^2=-q^2$ but we prefer the 
more general notation $t$ to avoid confusion with the incoming and outgoing 
photon virtualities.} and $\nu=(P\cdot K)/M$ as the 
independent variables. $M$ denotes the nucleon mass. They are related to the 
Mandelstam variables $s=(p+k)^2$ and $u=(p-k')^2$ through $s-u=4M\nu$ and 
$s+u+t=2M^2$. For convenience, we also introduce the usual polarization 
parameter $\varepsilon$ of the virtual photon, which can be related to the 
invariants $\nu$ and $t$ (neglecting the electron mass $m$):
\beqn
\varepsilon\,=\,\frac{\nu^2-M^2\tau(1+\tau)}{\nu^2+M^2\tau(1+\tau)},
\eeqn
with $\tau=-t/(4M^2)$. Elastic scattering of two spin $1/2$ particles is 
described by six independent amplitudes. Three of them do not flip the 
electron helicity \cite{marcguichon},
\beqn
T_{no\;flip}&=&\frac{e^2}{-t}
\bar{u}(k')\gamma_\mu u(k)\label{f1-3}
\\
&\cdot&
\bar{u}(p')
\left(\tilde{G}_M \gamma^\mu\,-\,
\tilde{F}_2\frac{P^\mu}{M}\,+\,
\tilde{F}_3\frac{\kdagger P^\mu}{M^2}\right)u(p),\nn
\eeqn
while the other three are electron helicity flipping and thus have in 
general the order of the electron mass $m_e$ \cite{jamarcguichon}:
\beqn
T_{flip}&=&\frac{m_e}{M}\frac{e^2}{-t}
\left[
\bar{u}(k')u(k)\cdot\bar{u}(p')\left(\tilde{F}_4\,+\,
\tilde{F}_5\frac{\kdagger}{M}\right)u(p)\right.\nn\\
&&\;\;\;\;\;\;\;\;\;\;\;+\;
\tilde{F}_6\bar{u}(k')\gamma_5u(k)\cdot\bar{u}(p')\gamma_5u(p)
\Big]
\label{f4-6}
\eeqn
\indent
In the one-photon exchange (OPE) approximation, two of the six amplitudes 
match with the electromagnetic form factors,
\beqn
\tilde{G}_M^{Born}(\nu,t)&=&G_M(t),\nn\\
\tilde{F}_2^{Born}(\nu,t)&=&F_2(t),\nn\\
\tilde{F}_{3,4,5,6}^{Born}(\nu,t)&=&0
\eeqn
where $G_M(t)$ and $F_2(t)$ are the magnetic and Pauli form factors, 
respectively. For further convenience we define also 
$\tilde{G}_E\,=\,\tilde{G}_M-(1+\tau)\tilde{F}_2$ and 
$\tilde{F}_1\,=\,\tilde{G}_M-\tilde{F}_2$ which in the Born approximation 
reduce to Sachs electric form factor $G_E$ and Dirac form factor $F_1$, 
respectively. It is useful to separate one- and two-photon exchange 
effects explicitly, 
\beqn
\tilde{G}_M&=&G_M+\delta\tilde{G}_M,\nn\\
\tilde{G}_E&=&G_E+\delta\tilde{G}_E,\nn\\
\tilde{F}_2&=&F_2+\delta\tilde{F}_2,
\eeqn
where $G_M$, $G_E$, and $F_2$ are the usual form factors, while the two photon 
effects are contained in the quantities $\delta\tilde{G}_M$, 
$\delta\tilde{G}_E$, $\delta\tilde{F}_2$. 
 
The unpolarized cross section is 
\beqn
\frac{d\sigma}{d\Omega_{Lab}}&=&\frac{\tau\sigma_R}{\varepsilon(1+\tau)}
\frac{d\sigma_0}{d\Omega_{Lab}},
\eeqn
with the usual Rutherford cross section 
\beqn
\frac{d\sigma_0}{d\Omega_{Lab}}&=&\frac{4\alpha^2\cos^2\frac{\Theta}{2}}{Q^4}
\frac{E'^3}{E},
\eeqn
$\Theta$ the electron Lab scattering angle and $E(E')$ the incoming (outgoing) 
electron Lab energy. The reduced cross section $\sigma_R$ is given by
\beqn
\sigma_R&=&G_M^2+\frac{\varepsilon}{\tau}G_E^2
+ 2G_M {\rm Re}
\left(\delta\tilde{G}_M+\varepsilon\frac{\nu}{M}\tilde{F}_3\right) \nn\\
&+&2\frac{\varepsilon}{\tau}G_E {\rm Re}
\left(\delta\tilde{G}_E+\frac{\nu}{M}\tilde{F}_3\right)+{\cal{O}}(e^4)
\eeqn

The OPE contributions to the form factors depend linearly on the lepton's 
charge, while the TPE contributions are quadratic in the lepton's charge. 
Therefore, the interference terms between the OPE and TPE will enter 
$e^-p$ and the $e^+p$ cross sections with opposite sign. 
The $e^+p/e^-p$ cross section ratio is thus given by
\beqn
R^{e^+/e^-}&\equiv&\frac{\sigma_{e^+p}}{\sigma_{e^-p}}\\
&=&1-\frac{4}{G_M^2+\frac{\varepsilon}{\tau}G_E^2}
\left[
G_M {\rm Re}
\left(\delta\tilde{G}_M+\varepsilon\frac{\nu}{M}\tilde{F}_3\right) \right.\nn\\
&&\;\;\;\;\;\;\;\;\;\;\;\;\;\;\;\;+\;\left.\frac{\varepsilon}{\tau}G_E {\rm Re}
\left(\delta\tilde{G}_E+\frac{\nu}{M}\tilde{F}_3\right)\right]\nn
\eeqn

For a beam polarized normal to the scattering plane, one can 
define a single spin asymmetry,
\beqn
B_n\,=\,
\frac{\sigma_\uparrow-\sigma_\downarrow}{\sigma_\uparrow+\sigma_\downarrow},
\eeqn
where $\sigma_\uparrow$ ($\sigma_\downarrow$) denotes the cross sesction for 
an unpolarized target and for an electron beam spin parallel (antiparallel) 
to the normal polarization vector defined as 
\beqn
S_n^\mu\,=\,
\left(0,\frac{[\vec{k}\times\vec{k'}]}{|\vec{k}\times\vec{k'}|}\right),
\eeqn
normalized to $(S\cdot S)=-1$. Similarly, one defines the target normal spin 
asymmetry $T_n$. It has been shown in the early 1970's \cite{derujula} that 
such asymmetries are directly related to the imaginary part of the $T$-matrix.
Since the electromagnetic form factors and the one-photon exchange amplitude
are purely real, $B_n$ obtains its finite contribution to leading order in the 
electromagnetic constant $\alpha_{em}$ from an interference between the Born 
amplitude and the imaginary part of the two-photon exchange amplitude.
In terms of the amplitudes of Eqs.(\ref{f1-3},\ref{f4-6}), the beam normal 
spin asymmetry is given by:
\beqn
B_n&=&-\frac{m_e}{M}\sqrt{2\varepsilon(1-\varepsilon)}\sqrt{1+\tau}
\left(\tau G_M^2+\varepsilon G_E^2\right)^{-1}\nn\\
&\cdot&
\left\{
\tau G_M {\rm Im}\tilde{F}_3\,+\,G_E {\rm Im}\tilde{F}_4
\,+\,F_1\frac{\nu}{M}{\rm Im}\tilde{F}_5
\right\}
\label{eq:bn_general}
\eeqn

\section{Imaginary part of the forward elastic $ep$-scattering amplitude}
\label{sec:im_2gamma}
\begin{figure}[h]
{\includegraphics[height=2cm]{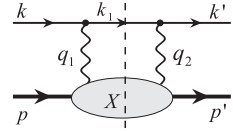}}
\caption{Imaginary part of the $2\gamma$-exchange diagram}
\label{fig:2gammadiag}
\end{figure}

The imaginary part of the diagram in Fig. \ref{fig:2gammadiag} is given by the 
integral
\beqn
{\rm Im}T_{2\gamma}&=&e^4\int\frac{d^3\vec{k}_1}{(2\pi)^32E_1}
\frac{1}{q_1^2q_2^2}l_{\mu\nu}\cdot{\rm Im}W^{\mu\nu},
\eeqn
where the leptonic tensor is given by
\beqn
l_{\mu\nu}&=&\bar{u}(k')\gamma_\nu(\keldagger_1+m_e)\gamma_\mu u(k).
\eeqn

The imaginary part of the spin-averaged part of the hadronic tensor is 
expressed in terms of the 
DIS structure functions $W_1$ and $W_2$. We will make use of Callan-Gross 
relation between them, thus we obtain 
\beqn
{\rm Im}W^{\mu\nu}&=&2\pi W_1
\left\{-g^{\mu\nu}\,+\,\frac{P^\mu q_1^\nu+P^\nu q_2^\mu}{P\tilde{K}}\right.
\nn\\
&&\left.\;\;\;\;\;\;
-\;\frac{(q_1\cdot q_2)}{(P\cdot\tilde{K})^2}P^\mu P^\nu\right\},
\eeqn
with $q_1^\mu=k-k_1$ the incoming and $q_2^\mu=k'-k_1$ the outgoing photon 
momenta and their average $\tilde{K}=\frac{q_1+q_2}{2}$.

We next contract the leptonic and hadronic tensors and keep the lepton mass 
non-zero in order to cross-check the calculation with the previous work for 
the beam normal spin asymmetry \cite{jaqrcs}. The result reads
\beqn
l_{\mu\nu}\cdot{\rm Im}W^{\mu\nu}&=&2\pi W_1
\left\{
m_e\bar{u}'u\frac{2Pk_1}{PK}
\right.\nn\\
&&+\bar{u}'\pgdagger u
\left[\frac{2K^2Pk_1}{(P\tilde{K})^2}
\left(\frac{Pk_1P\tilde{K}}{(PK)^2}+\frac{P\tilde{K}}{PK}-2\right)
\right.\nn\\
&&+\left.\left.\frac{Q^2}{P\tilde{K}}
\left(\frac{2PK}{P\tilde{K}}+\frac{P\tilde{K}}{PK}-2\right)
\right]\right\}
\label{eq:tensorcontr}
\eeqn

In the above equation, we used the notation $Q^2=-q_1^2=-q_2^2$. 
Throughout the calculation, we neglect terms 
$\sim\frac{P^2K^2}{(PK)^2}\approx\frac{1-\epsilon}{1+\epsilon}$ which are small 
in forward kinematics.

The above formula represents the TPE amplitude in the nucleon 
helicity-conserving channel, averaged over the spin projections. The dependence 
on the electron spins is retained. Our goal is to obtain the expressions for 
the invariant $ep$-scattering amplitudes defined in Eqs. 
(\ref{f1-3}, \ref{f4-6}). For this, we will need to restore the dependence on 
the nucleon spin. It can be done as follows. 
There exist four independent scalars which can be formed from the initial and 
final nucleon spinors with the $\gamma$-matrices and the four-vectors that fix 
the external kinematics. They are 
$\bar{N}'N$, $\bar{N}'\kdagger N$, $\bar{N}'\gamma_5\kdagger N$ and 
$\bar{N}'\gamma_5N$.\footnote{The vector $\bar{N}'\gamma^\mu N$, the  
axial vector $\bar{N}'\gamma^\mu\gamma_5 N$ terms, as well as the tensorial 
terms $\bar{N}'\sigma^{\mu\nu} N$ and $\bar{N}'\sigma^{\mu\nu}\gamma_5 N$ 
in the adopted formalism 
have to be contracted with the four independent four-vectors 
which form a basis, that is 
$K^\mu,P'^\mu=P^\mu-\frac{PK}{K^2}K^\mu,q^\mu,
N^\mu=\varepsilon^{\mu\alpha\beta\gamma}P'_\alpha K_\beta q_\gamma$. 
Working out the Dirac algebra, it can be shown that the only independent 
structures are $\bar{N}'N$, $\bar{N}'\kdagger N$, $\bar{N}'\gamma_5\kdagger N$ 
and $\bar{N}'\gamma_5N$}
However, the fourth structure vanishes 
in the exact forward limit and in the helicity conserving channel. 
Furthermore, the third structure drops after averaging over spins (that is, 
it is non-zero in the GDH sum rule-like cross section difference  
$T_{1/2,\lambda';1/2,\lambda}-T_{-1/2,\lambda';-1/2,\lambda}$, but not in the 
sum). 
Finally, we are left with only two structures $\bar{N}'N$, 
$\bar{N}'\kdagger N$. By a direct calculation, it can be shown that in the 
forward limit,
\beqn
1\,=\,\frac{1}{2}\sum_{spins}\left\{-\frac{M}{s-M^2}\bar{N}'N
+\frac{s+M^2}{(s-M^2)^2}\bar{N}'\kdagger N\right\}
\eeqn
\indent
Therefore, we now modify the hadronic tensor by substituting
\beqn
W_1\rightarrow W_1
\left\{-\frac{M}{s-M^2}\bar{N}'N
+\frac{s+M^2}{(s-M^2)^2}\bar{N}'\kdagger N\right\}
\label{eq:w1sub}
\eeqn

This procedure allows one to restore the nucleon spin dependence of the 
hadronic 
amplitude from its spin-averaged part in a correct way, as long as only 
cross section channel is used as the phenomenological input. 


\indent
We can now identify the invariant amplitudes $F_i$, $i=1,\dots,6$ comparing 
Eqs. (\ref{f1-3},\ref{f4-6},\ref{eq:tensorcontr},\ref{eq:w1sub}):
\beqn
{\rm Im}\tilde{G}_M&=&0\\
{\rm Im}\tilde{F}_2&=&-2\pi e^2t\frac{M^2}{s-M^2}\int_A
\left\{W_1(w^2,q_1^2,q_2^2)\right\}\nn\\
{\rm Im}\tilde{F}_3&=&\frac{s+M^2}{s-M^2}{\rm Im}F_2\nn\\
{\rm Im}\tilde{F}_4&=&
2\pi e^2t\frac{M^2}{s-M^2}\int_BW_1(w^2,q_1^2,q_2^2)\nn\\
{\rm Im}\tilde{F}_5&=&-\frac{s+M^2}{s-M^2}{\rm Im}F_4\nn\\
{\rm Im}\tilde{F}_6&=&0,\nn
\eeqn
where we introduced shorthands:
\beqn
&&\int_A\left\{W_1(w^2,Q_1^2,Q_2^2)\right\}
\,=\,\int\frac{d^3\vec{k}_1}{(2\pi)^32E_1}
\frac{1}{Q_1^2Q_2^2}\nn\\
&&\times\left[\frac{2K^2Pk_1}{(P\tilde{K})^2}
\left(\frac{Pk_1P\tilde{K}}{(PK)^2}+\frac{P\tilde{K}}{PK}-2\right)
\right.\nn\\
&&+\left.\frac{Q^2}{P\tilde{K}}
\left(\frac{2PK}{P\tilde{K}}+\frac{P\tilde{K}}{PK}-2\right)
\right]\,W_1(w^2,Q_1^2,Q_2^2)\nn\\
&&\int_BW_1(w^2,Q_1^2,Q_2^2)
\,=\,\int\frac{d^3\vec{k}_1}{(2\pi)^32E_1}
\frac{1}{Q_1^2Q_2^2}\nn\\
&&\times\frac{2Pk_1}{PK}W_1(w^2,Q_1^2,Q_2^2)
\label{eq:iAiB}
\eeqn

Note that the fact that Im$G_M$ vanishes implies that Im$F_1=-$Im$F_2$.
In the above formula, the first argument of $W_1$ is the invariant mass 
squared of the $\gamma^*p$ system, $w^2=(P+\tilde{K})^2$. In the integral 
$\int_A$, we have to keep the term $\sim K^2=-\frac{t}{4}$ till the end, 
since it is needed to cancel the $1/t$ behaviour of the integral $I_1$ 
introduced below.

\indent
The next step is to perform the phase space integrals in Eq. (\ref{eq:iAiB}). 
First, we express the nucleon structure function $W_1$ through the virtual 
photon cross section $\sigma_{\gamma^*p}(w^2,Q^2)$:
\beqn
W_1\,=\,\frac{w^2-M^2}{2\pi e^2}\sigma_{\gamma^*p}(w^2,Q^2).
\eeqn

\indent
We will need an explicit form of energy and $Q^2$ dependence of this cross 
section. We will use the phenomenological form proposed in Ref.\cite{cvetic} 
which amounts in factorization of the virtual photon cross section into 
purely $w^2$-dependent part, and the low-$x$ scaling variable 
$\eta$-dependent part:
\beqn
\sigma_{\gamma^*p}(w^2,Q^2)\,=\,\sigma_{\gamma p}^{Regge}(w^2)
\frac{I[\eta(\Lambda^2(w^2)),\eta_0]}{I[\eta_0,\eta_0]},
\eeqn
where $\sigma_{\gamma p}^{Regge}(w^2)$ is the total photoabsorption cross 
section in Regge regime and is conveniently parametrized in terms of two 
Regge trajectories,
\beqn
\sigma_{\gamma p}^{Regge}(w^2)\,=\,A_\rho \left(w^2\right)^{\alpha_\rho-1}
\,+\,A_P\left(w^2\right)^{\alpha_P-1}
\eeqn
with $A_\rho=(145\pm2)\mu b$, $A_P=(63.5\pm0.9)\mu b$, $\alpha_\rho=0.5$ and 
$\alpha_P=1.097\pm0.002$ the parameters of the $\rho$ and pomeron trajectories, 
respectively. In the above formula, $w$ should be taken in GeV. 
The scaling variable $\eta$ is defined as
\beqn
\eta\,=\,\frac{Q^2+m_0^2}{\Lambda^2(w^2)},
\eeqn
with its minimal value $\eta_0\equiv\eta(Q^2=0)=\frac{m_0^2}{\Lambda^2(w^2)}$.
Furthermore, the function of energy $\Lambda(w^2)$ is determined by a fit to 
DIS data which are well reproduced by 
\beqn
\Lambda^2(w^2)\,=\,C_1(w^2+w_0^2)^{C_2},
\eeqn
with $C_1=0.34\pm0.05$, $C_2=0.27\pm0.01$, $w_0^2=882\pm246$ GeV$^2$, and 
$m_0^2=0.16\pm0.01$ GeV$^2$. 
The general form of the function $I(\eta,\eta_0)$ can be found in 
\cite{cvetic}. Here, we will use its asymptotic form for low values of $\eta$,
\beqn
I(\eta,\eta_0)_{\eta\ll}\,=\,\ln\left(\frac{1}{\eta}\right).
\eeqn

Furthermore, we 
divide the virtual photon cross section in two parts, the $Q^2$ independent 
and the rest,
\beqn
\sigma_{\gamma^*p}(w^2,Q^2)&=&\sigma_{\gamma p}^{Regge}(w^2)\nn\\
&+&
(\sigma_{\gamma^*p}(w^2,Q^2)-\sigma_{\gamma p}^{Regge}(w^2))\nn\\
&=&\sigma_{\gamma p}^{Regge}(w^2)\nn\\
&-&\sigma_{\gamma p}^{Regge}(w^2)
\frac{\ln\left(1+Q^2/m_0^2\right)}{\ln\left(\frac{\Lambda^2}{m_0^2}\right)}
\eeqn

\subsection{Electron helicity-flip amplitudes}
We start with the integral $\int_B$. 
Introducing the dimensionsless variable $z=\frac{E_1}{E}$ and consistently 
neglecting terms $\sim t/s$, we can write the integral $\int_B$ as 
\beqn
&&\int_BW_1(w^2,q_1^2,q_2^2)\,=\,
\frac{(s-M^2)E^2}{16\pi^4e^2}\\
&&\times\int_{\frac{m_e}{E}}^{\frac{E_M}{E}}
dz\,z^2(1-z)\,\sigma_{\gamma p}(M^2+(s-M^2)(1-z))\nn\\
&&\times
\int\frac{d\Omega_1}{Q_1^2Q_2^2}
\left[1-
\frac{\ln\left(1+Q^2/m_0^2\right)}{\ln\left(\frac{\Lambda^2}{m_0^2}\right)}
\right]\nn
\eeqn

The integral corresponding to the first term in the square brakets 
was performed before (see \cite{javvcs},\cite{jaqrcs}). 
In order to obtain the correct $t$-dependence, we keep the photon indices in 
the denominator: then, the first term in the square braket leads to the 
leading $1/t$ behaviour. For the second term, the denominator can be written 
as $1/(Q^2)$, as this corresponds to neglecting the order $t^2$ 
corrections to the leading term. 
The second term can be related to the table integral
$\int\frac{\ln(1+x)}{x^2}dx\;=\;\ln x\,-\,\frac{1+x}{x}\ln(1+x)$. 
The result for the electron helicity-flip amplitudes then reads
\beqn
&&{\rm Im}\tilde{F}_4\;=\;
-\frac{M^2}{4\pi^2}\\
&&\times\int_{\frac{m_e}{E}}^{\frac{E_M}{E}}
dz\,(1-z)\,\sigma_{\gamma p}(w^2)
\left\{
\ln\left(\frac{z^2}{(1-z)^2}\frac{-t}{m^2}\right)\right.\nn\\
&&\left.+\frac{zt}{2m_0^2\ln\left(\frac{\Lambda^2}{m_0^2}\right)}
\left[\ln\frac{2Ez}{m_e}+\frac{1}{2}-\ln\left(1+\frac{4E^2z}{m_0^2}\right)
\right]\right\}\nn\\
&&{\rm Im}\tilde{F}_5\;=\;-\frac{s+M^2}{s-M^2}{\rm Im}F_4
\eeqn

Putting these expressions into Eq. (\ref{eq:bn_general}), we obtain our 
result for the beam normal spin asymmetry,
\beqn
\label{eq:bn_result}
&&B_n\;=\;
-\frac{m_e\sqrt{-t}}{4\pi^2}\frac{F_1(t)}{F_1^2(t)+\tau F_2^2(t)}\\
&&\times\int_{\frac{m_e}{E}}^{\frac{E_M}{E}}
dz\,(1-z)\,\sigma_{\gamma p}(w^2)
\left\{
\ln\left(\frac{z^2}{(1-z)^2}\frac{-t}{m^2}\right)\right.\nn\\
&&\left.+\frac{zt}{2m_0^2\ln\left(\frac{\Lambda^2}{m_0^2}\right)}
\left[\ln\frac{2Ez}{m_e}+\frac{1}{2}-\ln\left(1+\frac{4E^2z}{m_0^2}\right)
\right]\right\}\nn
\eeqn

\begin{figure}[ht]
{\includegraphics[height=7cm]{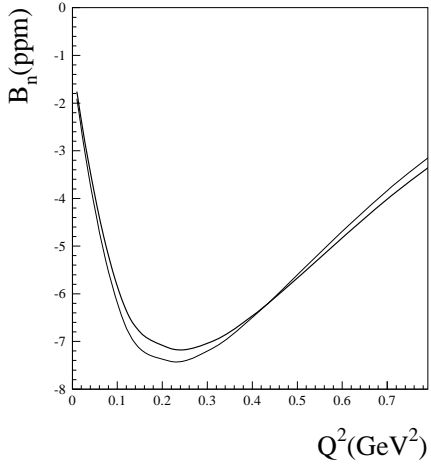}}
\caption{Beam normal spin asymmetry in the forward regime for the 
lab beam energy $E_lab=3$ GeV as function of $Q^2$. The thick line corresponds 
to the leading $t$ contribution, while the thin line corresponds to the full 
calculation with subleading terms in $t$.}
\label{fig:bn}
\end{figure}

Before presenting the results of 
the numerical integration, we recall that 
in the previous work, the cross section was assumed to be independent of the 
photon virtuality and roughly constant as function of $w^2$, 
$\sigma_{\gamma^* p}(w^2,Q^2)\approx\sigma_T=const.$ In that case, 
the expression for $B_n$ was obtained \cite{javvcs}:
\beqn
B_n&=&-\frac{m_e\sqrt{-t}\sigma_T}{4\pi^2}
\frac{F_1(t)}{F_1^2(t)+\tau F_2^2(t)}
\left(\ln\frac{\sqrt{-t}}{m_e}-1\right)\nn\\
&&\label{eq:bn_old}
\eeqn
\indent
In Fig. \ref{fig:bn}, we compare the results for the beam normal spin asymmetry 
of Eq. (\ref{eq:bn_result}) with the approximative formula of 
Eq. (\ref{eq:bn_old}). Comparing the thick and the thin lines, we see that the 
numerical impact of the subleading terms in $t$ is very small, and the 
leading $t$ approximation indeed dominates. 

\subsection{Electron helicity-conserving amplitudes}
We next turn to the calculation of the imaginary parts of the electron helicity 
conserving amplitudes $F_2$ and $F_3$. 
This amounts in calculating the electron phase space integral $\int_A$. 
As in the case of $\int_B$, we start with the angular integration. 
There are four independent integrals over the solid angle in terms of which 
all other integrals can be expressed. They are
\beqn
I_1&=&\int\frac{d\Omega_1}{Q_1^2Q_2^2}
\left[1-
\frac{\ln\left(1+Q^2/m_0^2\right)}{\ln\left(\frac{\Lambda^2}{m_0^2}\right)}
\right],\label{eq:i1-4}\\
I_2&=&\int d\Omega_1\frac{1}{Q^2}
\left[1-
\frac{\ln\left(1+Q^2/m_0^2\right)}{\ln\left(\frac{\Lambda^2}{m_0^2}\right)}
\right],\nn\\
I_3&=&\int\frac{d\Omega_1}{w^2-M^2+Q^2}
\left[1-
\frac{\ln\left(1+Q^2/m_0^2\right)}{\ln\left(\frac{\Lambda^2}{m_0^2}\right)}
\right],\nn\\
I_4&=&\int\frac{d\Omega_1}{(w^2-M^2+Q^2)^2}
\left[1-
\frac{\ln\left(1+Q^2/m_0^2\right)}{\ln\left(\frac{\Lambda^2}{m_0^2}\right)}
\right]\nn.
\eeqn
\indent
The integrals $I_3$ and $I_4$ do not contain any singularity, while one 
has to be careful with the first two integrals in Eq. (\ref{eq:i1-4}). 
In the limit of small electron mass, one has for the first terms in the 
square brakets (for details, see \cite{javvcs, jaqrcs})
\beqn
I_1&\approx&
\frac{2\pi}{-tE_1^2}
\ln\left(\frac{E_1^2}{(E-E_1)^2}\frac{-t}{m_e^2}\right),\nn\\
I_2&\approx&
\frac{\pi}{EE_1}\ln\frac{4E_1^2}{m_e^2}.
\eeqn
\indent
One can notice that in the limit of zero mass of the electron the first two 
integrals become singular. In the case of the beam asymmetry $B_n$, this 
singularity is cancelled by the overall factor $m_e$, 
thus leading to the dominant term $\sim m_e\ln\frac{-t}{m_e^2}$. 
In the case 
of helicity conserving amplitudes, no such singularity may occur since it does 
not contain the overall factor $m_e$ which would cancel the chiral singularity. 
Therefore, the divergent parts of the integrals $I_1$ and $I_2$ should cancel 
exactly in the integral $I_A$. 
The integral $\int_A$ can be written in terms of these four integrals as 
\beqn
&&\int_A\left\{W_1(w^2,0,0)\right\}
\,=\,\int\frac{E_1dE_1}{16\pi^3}
\frac{w^2-M^2}{2\pi e^2}\sigma_{\gamma p}(w^2)\nn\\
&&\times
\left\{\frac{4}{w^2-M^2}\frac{E^2+E_1^2}{2E(E-E_1)}
\left[I_2-2K^2\frac{E_1}{E}I_1\right]\right.\nn\\
&&\left.-\frac{4E}{E-E_1}\left[I_4+\frac{1}{w^2-M^2}\frac{E_1}{E}I_3\right]
\right\}.
\eeqn
\indent
Indeed, in the combination 
$I_2-2K^2\frac{E_1}{E}I_1$, the dependence on $m_e$ cancels,
\beqn
I_2-2K^2\frac{E_1}{E}I_1&=&\frac{\pi}{EE_1}\ln\frac{4(E-E_1)^2}{Q^2}.
\eeqn
\indent
Combining all the terms, we obtain:
\beqn
&&\int_A\left\{W_1(w^2,q_1^2,q_2^2)\right\}\nn\\
&&=\,\frac{1}{8\pi^3e^2}
\int_{\frac{m_e}{E}}^{\frac{E_M}{E}}
\frac{dz}{1-z}\sigma_{\gamma p}(w^2)\nn\\
&&\times
\left\{
(1+z^2)\ln\frac{2E}{Q}+(1+z+z^2)\ln(1-z)\right.\nn\\
&&-\frac{2E}{\sqrt{s}}
\frac{z}{1-\frac{M^2}{s}z}-z\ln\left(1-\frac{M^2}{s}z\right)\nn\\
&&+\frac{1}{\ln\frac{\Lambda^2}{m_0^2}}
\left[
\frac{1+z^2}{2}Sp\left(-\frac{4E^2}{m_0^2}z\right)\right.\nn\\
&&\;\;\;+\left[1+z\ln\left(1+\frac{4E^2}{m_0^2}z\right)\right]
\ln\frac{1-\frac{M^2}{s}z}{1-z}\nn\\
&&\;\;\;-\frac{2E}{\sqrt{s}}\frac{z}{1-\frac{M^2}{s}z}
\ln\left(1+\frac{4E^2}{m_0^2}z\right)\nn\\
&&\left.\left.\;\;\;+zSp\left(\frac{2E}{\sqrt{s}}\frac{z}{1-z}\right)
\right]
\right\}
\eeqn
\indent
At high energies, the upper limit of the integration goes as 
${\frac{E_M}{E}}\approx1-\frac{m_\pi}{E_{lab}}\to1$. Consequently, 
the integrand has a rather singular behaviour for photon energies justs above 
the pion production threshold. This singularity is compensated by the 
vanishing of the cross section at pion production threshold, as it is 
dictated by unitarity. In the high energy approach used here, 
to ensure the correct threshold behaviour 
we multiply the cross section by a function which vanishes smoothly at the 
threshold and goes to 1 at higher energies, such that the high energy region 
is not affected. Here, we will adopt the threshold regulator of the following 
form:
\beqn
f_{thr}(z)\,=\,\sqrt{\frac{\frac{E_M}{E}-z}{1-z}},
\eeqn
which obeys the properties required above. An introduction of such a threshold 
regulator does, of course, lead to a certain model dependence in the low energy 
part of the integral. It is worthwhile to treat low photon energy region 
more accurately, for example, by using some phenomenological model for the 
resonant region and match it to the high energies contribution, and we leave 
this improvement to an upcoming work. 
For the moment, we concentrate on the high energies-motivated calculation. 
In this framework, the presented approach corresponds to accounting for 
the non-resonant background which originates from the 
quark-hadron duality picture. 
Since vanishing of this background function at the pion production 
threshold is required by unitarity, the introduced threshold function is 
dictated by physics, though not unique.

Finally, we can write down the analytical part of the result for Im$F_2$,
\beqn
{\rm Im}\tilde{F}_2
&=&\frac{M^2Q^2}{8\pi^2(s-M^2)}
\nn\\
&\times&
\int_0^{\frac{E_M}{E}}
\!\!\!\!\!\!dz
\frac{\sigma_{\gamma p}(M^2+(s-M^2)(1-z))}{1-z}
\sqrt{\frac{\frac{E_M}{E}-z}{1-z}}\nn\\
&&\times
\left\{
(1+z^2)\ln\frac{2E}{Q}+(1+z+z^2)\ln(1-z)\right.\nn\\
&&-\frac{2E}{\sqrt{s}}
\frac{z}{1-\frac{M^2}{s}z}-z\ln\left(1-\frac{M^2}{s}z\right)\nn\\
&&+\frac{1}{\ln\frac{\Lambda^2}{m_0^2}}
\left[
\frac{1+z^2}{2}Sp\left(-\frac{4E^2}{m_0^2}z\right)\right.\nn\\
&&\;\;\;+\left[1+z\ln\left(1+\frac{4E^2}{m_0^2}z\right)\right]
\ln\frac{1-\frac{M^2}{s}z}{1-z}\nn\\
&&\;\;\;-\frac{2E}{\sqrt{s}}\frac{z}{1-\frac{M^2}{s}z}
\ln\left(1+\frac{4E^2}{m_0^2}z\right)\nn\\
&&\left.\left.\;\;\;+zSp\left(\frac{2E}{\sqrt{s}}\frac{z}{1-z}\right)
\right]
\right\}
\label{eq:imf2}\\
{\rm Im}\tilde{F}_3&=&\frac{s+M^2}{s-M^2}{\rm Im}\tilde{F}_2
\label{eq:imf3}
\eeqn

\section{Dispersive calculation for real parts of $F_{1,2,3}$}
\label{sec:re_2gamma}
In the previous section we obtained imaginary parts of the amplitudes 
$F_1$, $F_2$ and $F_3$. Real parts of these amplitudes enter the expression 
of the elastic $ep$ cross section. To calculate these, we use dispersion 
relations at fixed $t=-Q^2$. Dispersion relations capitalize on the idea of 
{\it analyticity} of the scattering amplitude, which states that the same 
scattering amplitude describe the reaction in all three distinct channels 
related by partial $CP$-transformation, the so-called {\it crossing}: 
$s$-channel $e(k)+N(p)\to e(k')+N(p')$, 
$u$-channel $e(-k')+N(p)\to e(-k)+N(p')$, and $t$-channel 
$e(k)+e(-k')\to N(-p)+N(p')$. For elastic $eN$ scattering, the invariant 
amplitudes possess definite symmetry properties between $s$- and 
$u$-channels. These properties become more transparent if introducing the 
explicitly 
crossing antisymmetric variable $\nu=\frac{s-u}{4M}$ which is positive in the 
$s$-channel kinematics and negative in the $u$-channel. At zero momentum 
transfer this variables reduces to the $lab$ energy of the electron. 
It can be shown that the amplitudes $\tilde{F}_{1,2}$ are odd functions of 
$\nu$, while $\tilde{F}_{3}$ is an even function. 
As a result, these amplitudes obey the dispersion relations of two different 
forms,
\beqn
{\rm Re}\tilde{F}_{1,2}(\nu,Q^2)&=&\frac{2\nu}{\pi}
{\cal{P}}\int\frac{d\nu'}{\nu'^2-\nu^2}{\rm Im}_s\tilde{F}_{1,2}(\nu',Q^2),
\nn\\
{\rm Re}\tilde{F}_{3}(\nu,Q^2)&=&\frac{2}{\pi}
{\cal{P}}\int\frac{\nu'd\nu'}{\nu'^2-\nu^2}{\rm Im}_s\tilde{F}_{3}(\nu',Q^2),
\label{eq:f123_dr}
\eeqn
where ${\rm Im}_s$ indicates that imaginary parts are calculated in the 
$s$-channel. We next discuss the convergence of these dispersion integrals. 
As can be easily derived from Eq. (\ref{f1-3}), at very high energies,
\beqn
{\rm Im}_s\tilde{F}_{1,2}(\nu\to\infty,Q^2)&\sim&\frac{1}{\nu}
T_{\lambda'\lambda'_N;\lambda\lambda_N},\nn\\
{\rm Im}_s\tilde{F}_{3}(\nu\to\infty,Q^2)&\sim&\frac{1}{\nu^2}
T_{\lambda'\lambda'_N;\lambda\lambda_N},
\eeqn
where $T_{\lambda'\lambda'_N;\lambda\lambda_N}$ stands for the helicity 
amplitudes for elastic $ep$-scattering with initial (final) electron (proton) 
helicities indicated. In Regge-regime, the asimptotic energy dependence of the 
latter in the helicity conserving channel is given by the pomeron trajectory, 
\beqn
T_{\lambda'\lambda'_N;\lambda\lambda_N}\,\sim\,\nu^{\alpha_P},
\eeqn
with $\alpha_P\approx1.08$. Inserting these asimptotics in the corresponding 
dispersion relations for the invariant amplitudes, we see that the dispersion 
integrals converge, and no subtractions are needed. We therefore can obtain 
clean predictions for the real parts of the invariant amplitudes from a 
dispersive calculation which capitalizes on the unitarity and analyticity of 
the invariant amplitudes.
\begin{figure}[h]
{\includegraphics[height=10cm]{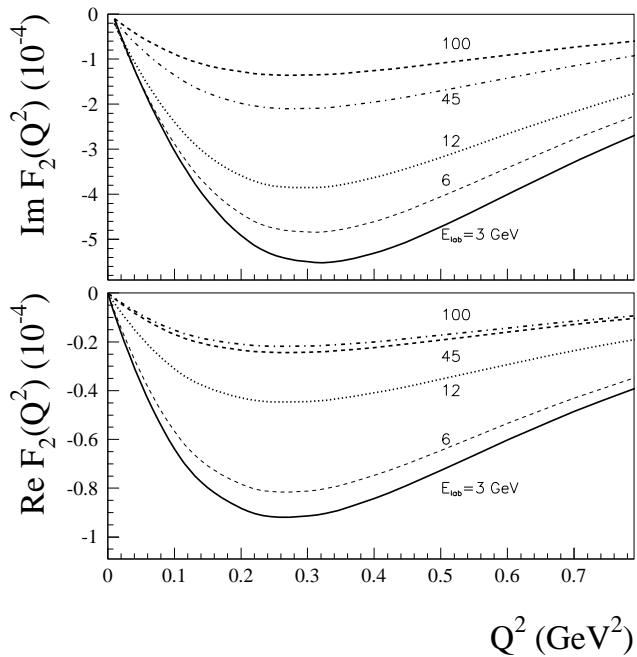}}
\caption{Imaginary (upper panel) and real (lower panel) part of the 
$2\gamma$-exchange contribution to the amplitude $\tilde{F}_2$.}
\label{fig:f2}
\end{figure}
\begin{figure}[h]
{\includegraphics[height=10cm]{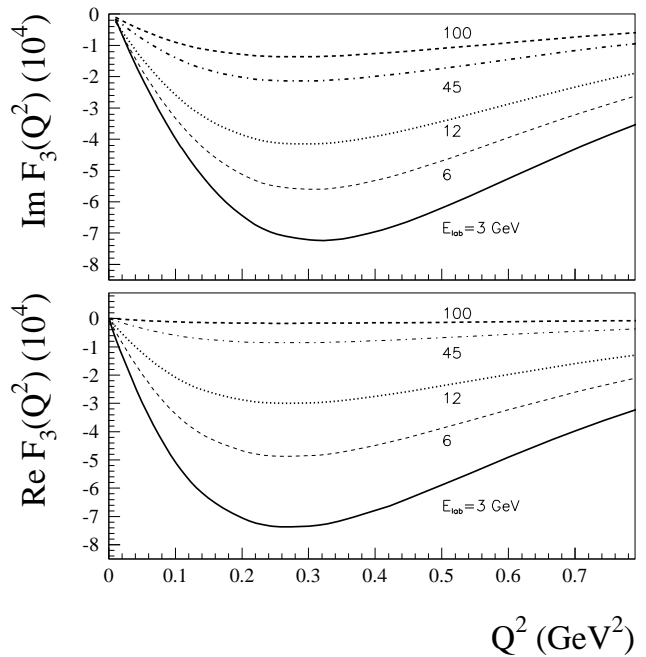}}
\caption{Imaginary (upper panel) and real (lower panel) part of the 
$2\gamma$-exchange contribution to the amplitude $\tilde{F}_3$.}
\label{fig:f3}
\end{figure}

\section{Results}
\label{sec:results}
We use a numerical routine for the principal value integrals, and 
present our results for the real and imaginary parts of the 
elastic $ep$-scattering amplitudes $F_2$ and $F_3$ for 5 different beam 
energies and as functions of the elastic momentum transfer $Q^2$. 
To present our results in the off-forward 
kinematics, we need phenomenological input in order to reproduce the Compton 
amplitude correctly at finite values of $t$. We use the exponential fit 
to Compton data, 
\beqn
\frac{d\sigma}{dt}\approx\left(\frac{d\sigma}{dt}\right)_{t=0}\times e^{Bt},
\eeqn
which provides a good description for $B\approx7$ GeV$^-2$ for 
values of $-t$ up to $\approx0.8$ GeV$^2$. Since $\frac{d\sigma}{dt}$ is 
related to the amplitude squared, while the total cross section to the 
amplitude's imaginary part, we conclude that \cite{afanas}
\beqn
\sigma_T(t)\,=\,\sigma_T(t_0)\times e^{\frac{Bt}{2}}.
\eeqn
In Figs. \ref{fig:f2} and \ref{fig:f3}, we display results for the amplitudes 
$\tilde{F}_2$ and $\tilde{F}_3$, respectively, in the forward regime using the 
model described above. The imaginary parts of the two amplitudes are related 
in a simple way, see Eq.(\ref{eq:imf3}), and therefore are roughly of the same 
size and of the same sign. However, Re$\tilde{F}_2$ is only about 10\% of 
Re$\tilde{F}_3$. 
This is due to the different forms of the dispersion relations which obey the 
two amplitudes. In the dispersion integral for $\tilde{F}_2$, 
${\cal{P}}\int\frac{d\nu'}{\nu'^2-\nu^2}{\rm Im}\tilde{F}_2$, 
the low values of $\nu'$, where the denominator is negative, 
have more impact than in that for $\tilde{F}_3$, thus we observe a substantial 
cancellation in this integral. 
Instead, for $\tilde{F}_3$, the extra power of the energy in the numerator, 
${\cal{P}}\int\frac{\nu'd\nu'}{\nu'^2-\nu^2}{\rm Im}\tilde{F}_3$
shifts the emphasis on higher values of energy and suppresses low energies, 
therefore the cancellation between the two regions is only very moderate.
\begin{figure}[h]
{\includegraphics[height=10cm]{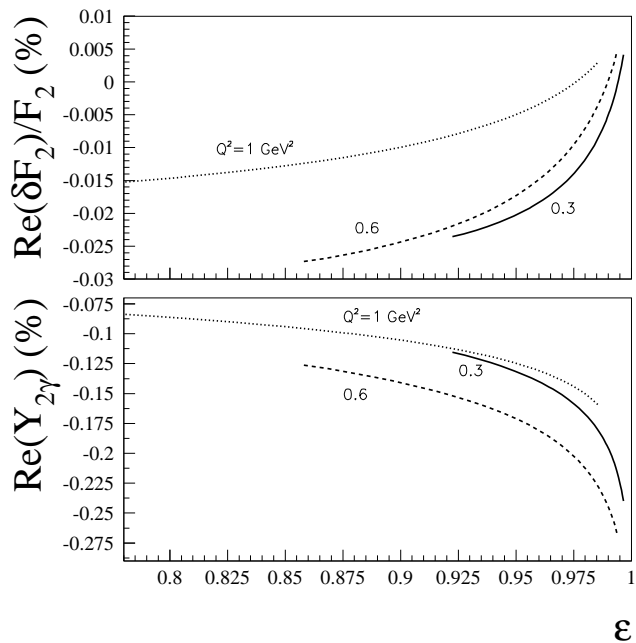}}
\caption{Real part of $\delta\tilde{F}_2$ normalized to the OPE value of $F_2$ 
at three different values of $Q^2$ and as function of $\epsilon$ (upper panel).
The same for the 
$2\gamma$-amplitude $Y_{2\gamma}$ (see text for the definition) is displayed 
in the lower panel.}
\label{fig:f23_epsilon}
\end{figure}
\begin{figure}[h]
{\includegraphics[height=10cm]{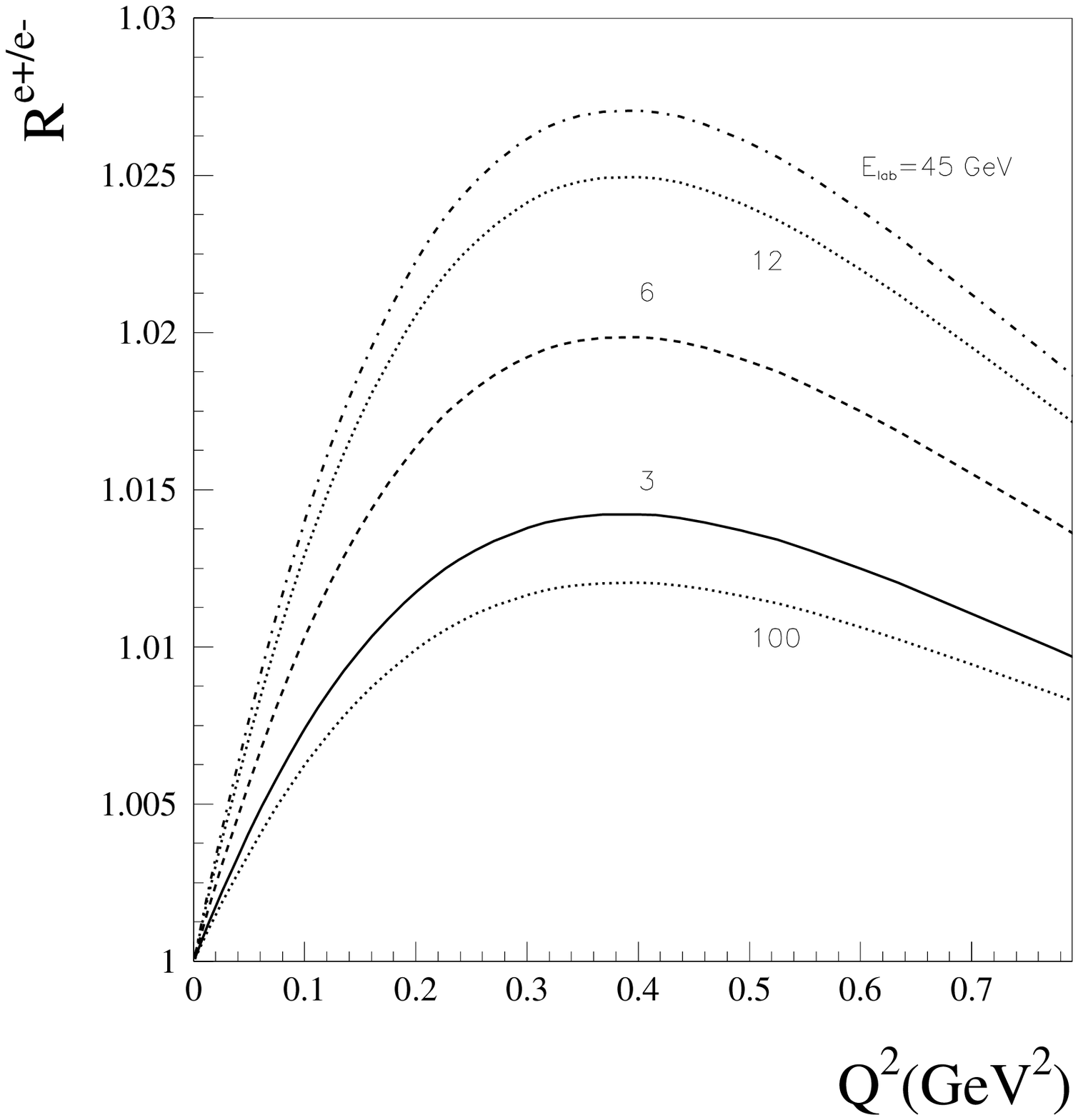}}
\caption{$e^+/e^-$ asymmetry in the forward regime for five different 
values of the lab beam energy as function of $Q^2$.}
\label{fig:epm}
\end{figure}
In Fig. \ref{fig:f23_epsilon}, we show the relative contribution of the 
TPE effects to the form factors weighted to the OPE values, 
$\frac{{\rm Re}\delta\tilde{F}_2}{F_2}$ and 
$Y_{2\gamma}=\frac{\nu}{M}\frac{{\rm Re}\tilde{F}_3}{G_M}$. 
It can be seen that the contribution to $F_2$ is fairly small. The 
combination $Y_{2\gamma}$ is quite small (0.1\%) at intermediate $\epsilon$'s, 
as well, but grows rapidly with $\epsilon$ approaching 1, i.e. 
at higher energies. In this plot,  we observe a very steep $\epsilon$ 
dependence of the generalized form factors at large $\epsilon$ values. As one 
goes to higher momentum transfers, though, this behaviour becomes smoother. 
At very high energies, the TPE contributions should decrease, according to 
the high energy asymtotics discussed above. 
Since we are interested in low values of $t$, such energies would correspond 
to the values of $\epsilon$ extremely close to 1 and cannot be displayed on 
this plot. 

Finally, we turn to the $e^+/e^-$ cross section ratio. 
In the forward kinematics and 
recalling that in forward regime $\delta \tilde{G}_M=0$, we obtain
\beqn
R^{e^+/e^-}=1-\frac{4}{F_1^2+\tau F_2^2}
{\rm Re}\left[F_1\frac{\nu}{M}\tilde{F}_3 - G_E \delta\tilde{F}_2\right]
\eeqn
The numerical results are shown in Fig. \ref{fig:epm}. 
In the forward regime, the positron cross section is expected to exceed that 
for electrons by a few procents: from 1.4\% for $E_{lab}=3$ GeV to 2.8\%  
for $E_{lab}=45$ GeV. If going to even higher beam energies, the ratio drops 
due to the high energy asymptotics of the invariant amplitudes. 
At the lab energies as high as several tens GeV, the dispersive contributions 
presented here are expected to be the dominant effect, while the elastic 
contribution (nucleon intermediate state in the blob in 
Fig. \ref{fig:2gammadiag}) should be relatively small. So, the above results 
should give a correct estimate of the TPE contribution to the 
elastic cross section in forward regiime.

\section{Conclusions}
We presented a dispersive calculation for the real parts of the electron 
helicity-conserving amplitudes $\tilde{F}_{1,2,3}$ for elastic 
$e^\pm p$-scattering in forward regime. Their imaginary parts in this regime 
can be related to the DIS structure functions. We used the phenomenological 
$w^2$ and $Q^2$ dependence of the structure functions. 
The imaginary parts of the $ep$ scattering amplitudes are IR finite, 
and we demonstrated that their high energy asymptotics ensure 
the convergence of the unsubtracted dispersive integrals for the 
corresponding imaginary parts. Due to differrent forms of the dispersion 
relations for the amplitudes $F_2$ and $F_3$, the resulting real parts have 
very different values: the dispersive effects contribute only about $0.05$\% 
of the low $Q^2$-value of $F_2$, while they account for a half to few per 
cents in the case of $Y_{2\gamma}=\frac{\nu}{M}\frac{{\rm Re}\tilde{F}_3}{G_M}$.
We predict the ratio of the $e^+p$ to $e^-p$ cross sections to be larger than 
1, and the predicted values for that ratio range from 1.4\% at $E_{lab}=3$ 
GeV to 2.8\% at $E_{lab}=45$ GeV. 

\acknowledgments
This work was supported under U.S. DOE contract DE-FG02-05ER41361.

\end{document}